\documentclass[11pt, a4paper, logo, twocolumn, external, copyright, nonumbering]{deepmind}
\usepackage[numbers]{natbib}
\usepackage{subcaption}
\usepackage{colortbl}

\title{Learned Force Fields Are Ready For Ground State Catalyst Discovery}

\correspondingauthor{mschaarschmidt@google.com, mriviere@deepmind.com, 2018jgodwin@gmail.com}

\reportnumber{} 

\author[1]{Michael Schaarschmidt}
\author[1]{Morgane Rivière}
\author[2]{Alex M. Ganose}
\author[1]{James S. Spencer}
\author[1]{Alexander L. Gaunt}
\author[1]{James Kirkpatrick}
\author[1, 3, 4]{Simon Axelrod}
\author[1]{Peter W. Battaglia}
\author[1]{Jonathan Godwin}

\affil[1]{DeepMind}
\affil[2]{Imperial College London}
\affil[3]{Harvard University}
\affil[4]{Massachusetts Institute of Technology}

\begin{abstract}
We present evidence that learned density functional theory (``DFT'') force fields are ready for ground state catalyst discovery. Our key finding is that relaxation using forces from a learned potential yields structures with similar or lower energy to those relaxed using the RPBE functional in over 50\% of evaluated systems, despite the fact that the predicted forces differ significantly from the ground truth. This has the surprising implication that learned potentials may be ready for replacing DFT in challenging catalytic systems such as those found in the Open Catalyst 2020 dataset. Furthermore, we show that a force field trained on a locally harmonic energy surface with the same minima as a target DFT energy is also able to find lower or similar energy structures in over 50\% of cases. This ``Easy Potential'' converges in fewer steps than a standard model trained on true energies and forces, which further accelerates calculations. Its success illustrates a key point: learned potentials can locate energy minima even when the model has high force errors. The main requirement for structure optimisation is simply that the learned potential has the correct minima. Since learned potentials are fast and scale linearly with system size, our results open the possibility of quickly finding ground states for large systems.
\end{abstract}

\begin{document}

\maketitle

\section{Introduction}

In July 2022 the single biggest workload of the largest UK supercomputer, ARCHER2 \cite{archer2}, was density functional theory  \cite{PhysRev.136.B864, PhysRev.140.A1133} calculations using the Vienna \textit{ab initio} Simulation Package (``VASP'') \cite{PhysRevB.47.558, KRESSE199615, PhysRevB.54.11169}. The hope of such research is that by screening a very large number of compounds for specific properties, we can find answers for the grand scientific challenges of our transition to a green economy: better batteries, solar cells and cheap renewable fuels. This project also has significant expense: ARCHER2 cost £79M to build. 

An important use case for DFT is structure optimisation. The goal of structure optimisation is to find the global minimum of the potential energy surface (PES) of a material, a quantity that helps characterize its functional properties. The success of the optimisation depends on the complexity of the PES, the optimisation algorithm and the DFT approximation known as the exchange-correlation functional.

\begin{figure*}[h]
\centering
\begin{minipage}{0.6\textwidth}
    \centering
    \includegraphics[width=0.9\textwidth]{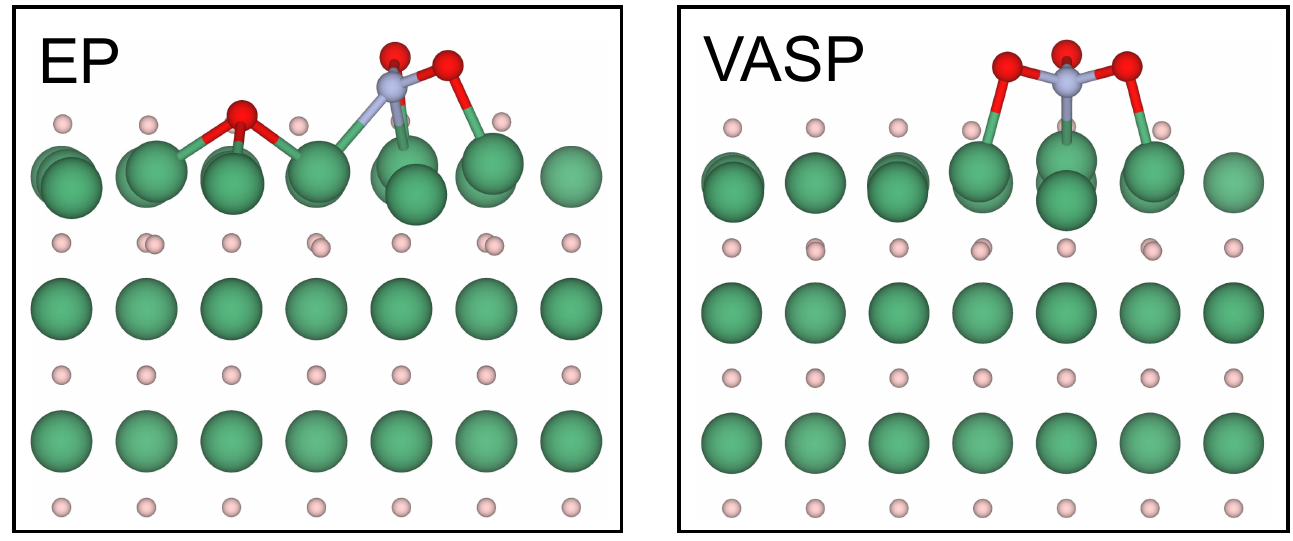}
\end{minipage}\hfill
\begin{minipage}{.4\textwidth}
    \centering
    \includegraphics[width=0.9\textwidth]{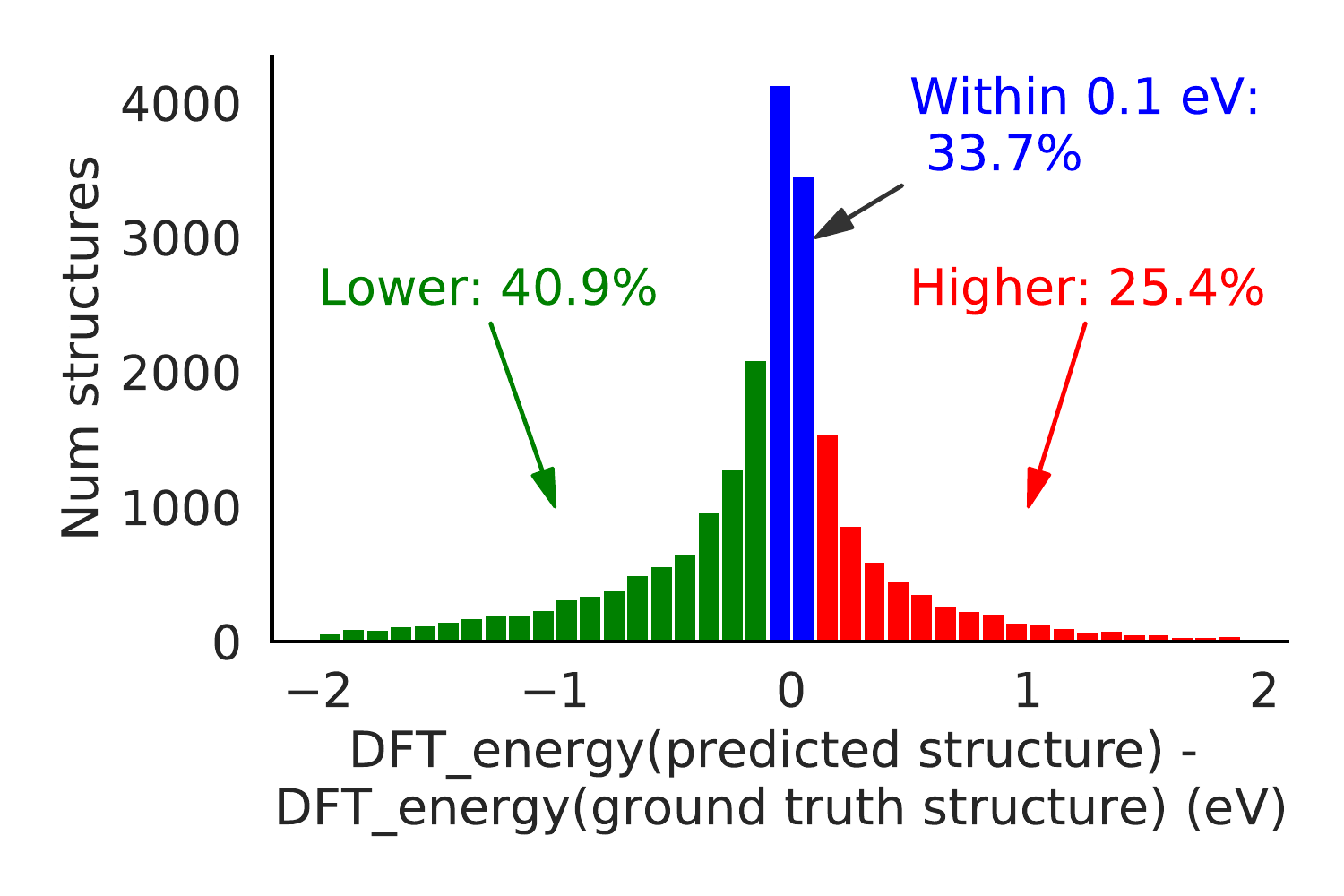}
\end{minipage}
\caption{\textbf{Summary and example of key findings.} ML-FF relaxations find lower energy structures on average than DFT relaxations with VASP. The structure found by relaxing ML-FF is of the same or lower energy in an out-of-distribution dataset for over 50\% of cases. Left: Example of an Easy Potential (EP) relaxed structure with lower energy (by 3.44 eV) than the VASP-relaxed geometry. The VASP result shows NO3 adsorbed on the surface of niobium hydride (NbH). In our ML calculation, the NO3 has reacted with the surface to form NO2 and O. This illustrates that our force fields are capable of describing chemical reactions, not just atomic relaxations. Right: Distribution of energy differences between ML-FF and DFT VASP optimization on OOD\_BOTH. \label{fig:energy_dist} }
\end{figure*}

Here we present the surprising result that machine learned potentials trained on DFT data consistently find low energy structures of the target functional outside of the training distribution. This occurs even when force estimates are inaccurate or noisy, and even when the model is trained on a simple harmonic approximation to the energy around ground states in the training set. Our key result is shown in Fig. \ref{fig:energy_dist}: in over 50\% of cases on out of distribution tasks, our learned potential finds structures with similar or lower energies to those optimised with RPBE in VASP using the same initial geometry. The median convergence time is one minute, which is two orders of magnitude faster than traditional DFT codes. This is despite being trained on a dataset where training labels do not represent true minima either.

Our findings have a surprising implication: learned force fields may already be ready for use in ground state discovery for catalytic systems. If DFT optimisation is a useful tool for catalyst discovery, and force field optimised structures have a lower mean energy than DFT geometries, then force fields are useful for catalyst discovery. This is despite the fact that learned models still have non-negligible errors with respect to DFT energies and forces.

Our experiments suggest that learned potentials succeed because they have the correct energy minima, even if their force errors are high. To test this hypothesis, we trained a force field on a local harmonic approximation to the DFT energy around optimised structures in the Open Catalyst 2020 ("OC20") dataset \cite{Chanussot2020TheOC}. We call this an ``Easy Potential'' (EP) because it is so simple. Potentials trained on actual DFT data are termed ML-FF, for ``machine-learned force fields''. Optimising structures with this Easy Potential, and then refining them with an ML-FF, yielded even lower energy structures on average than optimisation with an ML-FF alone. In addition to explaining the success of learned force fields, this experiment indicates that Easy Potentials can accelerate and improve optimisation. This further increases the utility of machine learning in catalyst discovery. 

Finally, when combining Easy Potentials with VASP relaxations in a hybrid approach, we find that the structures match or exceed the stability of VASP-optimized geometries in 88\% of cases. The mean improvement is 0.46 eV, which is remarkably large. Further, these relaxations require half as many DFT steps as VASP optimization from scratch. This means that even practitioners not ready to fully embrace learned force fields can dramatically improve and accelerate their computations by initialising DFT-based relaxations through Easy Potentials. 

\section{Results}
We are given a set of atoms  
\begin{align*}
S = \{(a_1, p_1), (a_2, p_2), \dots,  (a_{\left| S \right|}, p_{\left| S \right|}) \}
\end{align*}
specified by their positions $p_i \in \mathbb{R}^3$ and atomic numbers $a_i \in \{1, \dots, 118\}$. Denoting the set of atomic positions as $P = {p_i, \dots, p_{\left| S \right|}}$, the forces exerted on the atoms are defined as $\vec{F}=-\nabla_{\left\Vert P \right\Vert}E(S) \in R^{\left| S \right| \times 3}$, where $E(S)$ is the potential energy for $S$ and $\left\Vert \cdot \right\Vert$ is the concatenation operator. The ground state of a PES is the set of atomic positions that minimises the energy, $P_{min} = \text{min}_P(E(S))$.

\subsection{Density Functional Theory}
DFT is a method to compute the total energy of a system using a functional of the electron density \cite{PhysRev.140.A1133,PhysRevLett.77.3865,PhysRev.136.B864}. DFT exists along a spectrum of methods to calculate the energy and forces of an atomic configuration that trade off computation time and accuracy. On one end there are traditional force fields, which include simple classical physics-inspired terms like the Lennard-Jones potential \cite{jones1924determination}. These potentials can be rapidly evaluated, but often lack transferability and/or accuracy. On the other end, there are quantum chemical methods such as CCSD(T) \cite{RevModPhys.79.291} and quantum Monte Carlo \cite{RevModPhys.73.33,Motta2018-wm}, which are highly accurate but too expensive for all but the simplest systems. Reducing computational cost while maintaining accuracy is an ongoing effort \cite{guo2018communication}.

DFT is the most widely used method for computing energies for systems in condensed matter physics, materials science, and chemistry, with key papers having over 100 000 citations \cite{PhysRevLett.77.3865}. DFT's success comes from being sufficiently accurate for many applications, particularly in materials science and medicinal chemistry, but with an $O(N^3)$ run time and moderate prefactor that make it usable in practice. An important use case for DFT is structure optimisation, in which the atomic configuration with the minimum energy is found using a gradient-based optimisation algorithm. This process is also known as relaxation or finding the ground state of a material. The ground state energy of materials can be used to derive many thermodynamic properties, such as adsorption and formation energies \cite{Chanussot2020TheOC}. For this reason, structure optimisation using DFT is a workhorse of computational materials science.

Structure optimisation can be sensitive to the choice of functional and the choice of optimisation algorithm \cite{heydEfficientHybridDensity2004}. DFT codes such as VASP \cite{PhysRevB.47.558, KRESSE199615, PhysRevB.54.11169} and Quantum Espresso \cite{Giannozzi_2009} use conjugate gradients or quasi-Newton derived methods such as RMM-DIIS \cite{Wood_1985} and LBFGS \cite{Liu89onthe} for optimisation. These methods are tightly integrated into the software packages.

\subsection{Learned Interatomic Potentials}

Despite its favorable cost-accuracy trade-off relative to other quantum chemical methods, DFT is still rather expensive. Applications such as ab initio molecular dynamics (``MD'') \cite{dfrenkel96:mc}, which require thousands of energy calculations for each system, are often prohibitively expensive. Further, its cubic scaling makes DFT very challenging for evaluating large systems of many thousands of atoms, though linear scaling algorithms are under constant development \cite{prentice2020onetep}. The goal of learned atomic potentials \cite{unke2021machine} is to replace DFT with a neural network that is as accurate as DFT, but that is much faster to evaluate. 

Given a training set of set of atomic structures with associated DFT forces and energies $X = \{(s_1, E(s_1), \vec{F}_1),  \dots, (s_{|X|}, E(s_{|X|}), \vec{F}_{|x|}))\}$ we train a learned interatomic potential by minimising the approximation error of the energies and forces across a sample of DFT calculations, \\ $min_{\theta} \, \mathbb{E}_i(Error(E_{\theta}(s_i), E(s_i)) + Error(\vec{F}_{\theta}(s_i), \vec{F}(s_i)))$ where $\mathbb{E}$ is the expectation. Many learned interatomic potentials compute the forces by differentiating the energy with respect to the atomic positions, while others learn the forces independently \cite{Hu2021ForceNetAG, godwin2022simple}. We also predict forces directly, since this is faster than differentiating the energy.

A common class of neural networks used for learned atomic potentials are Message Passing Neural Networks \cite{Battaglia2018RelationalIB, Gilmer2017NeuralMP}, which represent a system as a spatial graph. Atoms are represented as nodes, with edges between atoms constructed based on inter-atomic distances and using a cutoff radius. Atom nodes and inter-atomic edges are then featurized using learned embeddings of atom types \cite{Schtt2017SchNetAC} as well as varying representations of positional information. In our work we use the Graph Net Simulator (GNS) \cite{pmlr-v119-sanchez-gonzalez20a}, a generic architecture suitable for molecular property prediction \cite{godwin2022simple}. Other architectures such as SchNet \cite{Schtt2017SchNetAC}, DimeNet \cite{2020DirectionalMP, 2020Dimenet++}, GemNet \cite{gemnet2021, gemnet-oc}, and PaiNN \cite{schutt2021equivariant} can also be used for the tasks in this work.

\begin{figure*}[ht!]
    \centering
    \includegraphics[width=\textwidth]{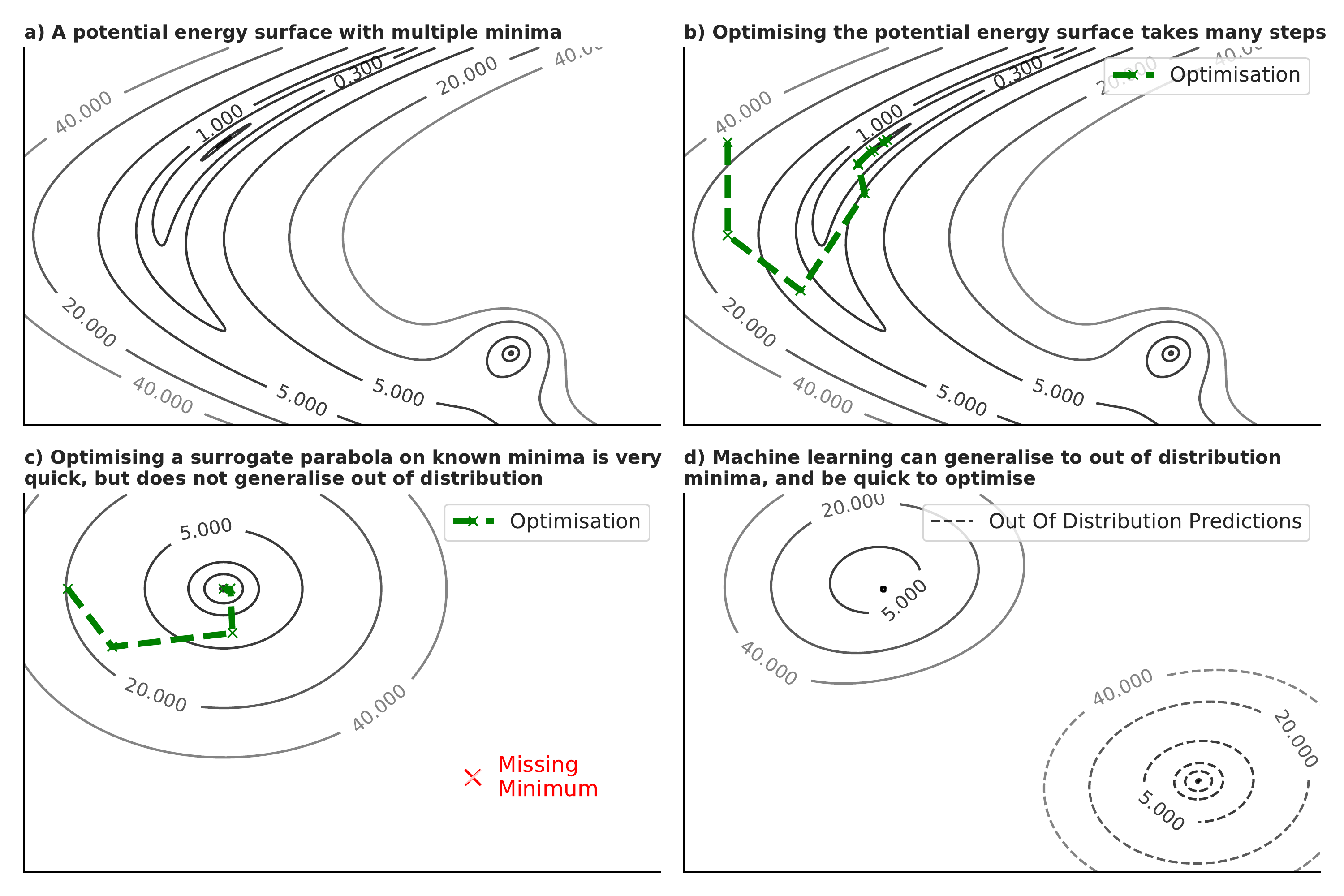}
    \caption{\textbf{Finding minima of difficult-to-optimise PESs using Easy Potentials.} a) A complicated PES with multiple minima. In b) we see that even simple 2D PESs can be challenging and slow to optimise. Parabolas as in c) centered on local minima are easy to optimise, but cannot be used to find new ground states or generalise to new systems. However, training to match such parabolas in similar systems can help generalise to new ground states. Machine learning potentials as in d) can have the best of both worlds: easy to optimise and generalise to new systems and new regions of phase space.}
    \label{fig:iso_figure}
\end{figure*}

Our learned potential, hereafter referred to as ML-FF, is trained and evaluated on different subsets of the OC20 dataset \cite{Chanussot2020TheOC}. OC20 is the largest dataset of ground state structures generated with the same DFT settings throughout, and is thus the most suitable for analysing ground state discovery. The S2EF (``structures to energy and forces'') subset has 134 million point-wise DFT calculations of both energies and forces. These are used to train models that predict $E$ and $\vec{F}$ for use in relaxation. Four validation datasets with approximately 25k structures each (initial and relaxed) are available to benchmark the model performance. The four validation sets have in-distribution catalysts and adsorbates (\textit{ID}), out of distribution catalysts (\textit{OOD\_CAT}), out of distribution adsorbates (\textit{OOD\_ADS}), and out of distribution catalysts and adsorbates (\textit{OOD\_BOTH}). Test sets are not publicly available and hence were not used in this work. We also note that the labels for the validation and test sets are in practice often wrong and do not correspond to the true minimum. Our main purpose in using the validation set labels is to determine when our model produces higher quality output than OC20 itself.

We train our model following the training method described by Zaidi et al. \cite{zaidi22} (full details in Methods). The model has a mean absolute error (MAE) of 0.029-0.038 eV/\AA \ for forces and 0.18-0.25 eV for energies, and 0.5 cosine similarity for forces on validation datasets. This error is similar to recent results in the literature; for example, GemNet \cite{gemnet2021} in 2021 reported 0.6 cosine similarity and 0.29 eV energy MAE on the OC20 test sets. We relax the structures using the Adamw optimiser \cite{xiao2018dynamical}, a weighted adaptive momentum optimiser that we found to best handle noisy force predictions.

\subsection{Easy Potentials}
While structure optimisation is typically guided by forces, one can also find ground states by following the gradient of a surrogate potential $\tilde{E}$ with the same global minimum as $E$. Indeed, one can choose a surrogate that is smoother than $E$, leading to faster optimisation and a higher likelihood of escaping local minima. 

\begin{table*}[ht!]
    \centering
    \begin{tabular}{c|ccc|>{\columncolor{lightgray}}c|cc}
          &  \multicolumn{3}{c}{Predicted vs. OC20 DFT energies} & Combined & Mean & Mean \\
     Dataset      & lower & within $0.1$ eV  & higher & lower/equal  &  ${\Delta} E$ (eV) & RMSD (\r{A})\\
    \midrule
     ID & 35.94\% & 37.19\% & 26.86\% & 73.13\% &  $-$0.07 &0.403  \\
     OOD\_CAT &  39.04\% & 33.03\% & 27.93\% & 72.07\% &  $-$0.11 & 0.403\\
     OOD\_ADS  & 37.19\% & 37.18\% & 25.62\% & 74.37\% & $-$0.05 & 0.402\\
     OOD\_BOTH  & 40.91\% & 33.69\% & 25.40\% & 74.6\% & $-$0.13 & 0.332\\

    \end{tabular}
    \caption{\textbf{Main ML-FF results on OC20 validation datasets.} For each of the validation sets, we compare DFT energies of ML-FF-relaxed structures with DFT energies of DFT-relaxed structures. We report the mean difference in relaxed DFT energies, given by (energy of ML-FF-relaxed structures $-$ energy of DFT-relaxed structures), i.e. lower is better.}
    \label{tab:result-overview}
\end{table*}

This point is illustrated in Fig. \ref{fig:iso_figure}. We start with two PESs, one a complicated energy surface (top left), and the other a parabola (bottom left) that has been centered on the minimum of the first surface. Optimising the parabola is fast and easy, taking only five steps of optimisation. The complicated PES is hard to optimise, taking 23 steps in total. However, constructing the parabola requires already knowing the locations of the local minima, and so it cannot be used for structure optimisation. This is where machine learning can help: when trained to match the parabola in different but related systems, the Easy Potential generalises to minima in the system of interest, while maintaining the same advantageous optimisation qualities (bottom right). 

The above example emphasises an important point: the only requirement for a learned potential being useful for material structure optimisation is that its forces are close to zero near the ground states of the target DFT functional. Indeed, our experiments show that Easy Potentials successfully \textit{identify low energy configurations outside of their training set even when they exhibit high force prediction error in high energy regions}. Counter-intuitively, this also means that ML-FF with high force prediction errors in areas outside a minimum may offer optimisation benefits, as noisy gradients \cite{neelakantan2017adding} can be a successful strategy to avoid getting stuck in local minima. %

For each ground state $\hat{s}$ in the training set, the Easy Potential approximates the energy of nearby states $s$ as $E(s) = E(\hat{s}) + \sum_i \frac{1}{2}||p_i - \hat{p}_i||_2^2$, where the sum is over each atom $i$ in $s$ with position $p_i$, and $|| \cdot ||_2$ is the L2 norm. The specific construction of the neural network predicting $E(s)$ in this manner is inspired by Godwin et al's work into regularisation to prevent over-smoothing \cite{godwin2022simple}. The forces are then $\Vec{F}_i(s) = \hat{p}_i - p_i$. Nearby states were sampled from intermediate points on the DFT relaxation trajectory, then augmented with additive Gaussian noise. Easy Potential gives rise to large energy and force errors when relaxed naively as a result of small errors in bond distances. As a result, we adopt a two stage relaxation procedure. First, we relax Easy Potential, and then we resolve bond errors with a second stage of relaxation using ML-FF. Adamw is again used as the optimiser. Full details of the relaxation and training procedures can be found in the Methods section.

\begin{table*}[t!]
    \centering
    \begin{tabular}{c|ccc|>{\columncolor{lightgray}}c|cc}
          &    \multicolumn{3}{c}{Predicted v. OC20 DFT energies}  & Combined  & Mean  & Mean\\
    Dataset &   lower & within $0.1$ eV margin & higher & lower/equal   & ${\Delta} E$ (eV) & RMSD (\r{A}) \\
    \midrule
    QE baseline  & & & & \\ 
     ID & 42.24\% & 35.40\% & 22.36\% & 77.64\% & $-$0.23 & 0.418\\
     OOD\_CAT & 45.54\% & 32.18\% & 22.28\% & 77.72\% & $-$0.07 & 0.416\\ 
     OOD\_ADS  &  42.27\% & 34.72\% & 23.00\% & 77.00\% & -0.18 & 0.434\\
     OOD\_BOTH  & 46.24\% & 31.85\% & 21.91\% & 78.09\% & $-$0.26 & 0.360 \\ 
     \\ 
     \midrule
     VASP sample & & & & \\ 
     (OOD\_BOTH)  & & & & \\      
     EP-default & 30.13\% & 28.42\% & 41.45\% & 58.55\% & $-$0.13 & 0.361\\
     EP-Extra & 33.80\% & 30.48\% & 35.72\% & 64.28\% & $-$0.23 & 0.363 \\
     EP+DFT & 44.22\% & 44.44\% & 11.34\% & 88.66\% & $-$0.46 & 0.355 \\
    \end{tabular}
    \caption{\textbf{Easy Potentials analysis.} A two-stage relaxation with Easy Potentials and ML-FFs improves results while using fewer relaxation steps. Quantum Espresso results on the full datasets are given above, and a sample of VASP calculations carried out on 1000 structures is given below. We also show a VASP comparison with additional budget in the refinement stage of EP ('EP-extra), as well as EP combined with DFT ('EP+DFT') relaxations, which both significantly improve results.}
    \label{tab:Easy Potential-performance}
\end{table*}

\subsection{ML-FF Relaxations}
Our first experiments demonstrate the ability of ML-FF relaxations to find minimum energy wells of the target DFT energy. Table \ref{tab:result-overview} compares the energies of relaxed ML-FF structures (``predicted structures'') with structures contained in the OC20 datasets. Since we have limited access to VASP, we obtain these results using DFT in Quantum Espresso for each set of structures. We use DFT settings as close as possible to those that generated the OC data. (As discussed below, VASP was also used on a small subset of the data to confirm that our results were not artifacts of Quantum Espresso.) The Quantum Espresso settings yield 93\%-96\% convergence rates on the validation datasets. Full details of the parameters used, and a comparison between VASP and Quantum Espresso can be found in the methods.

Table \ref{tab:result-overview} shows that, for each of the four validation sets containing 25 000 structures each, the mean energy of predicted structures is lower than that of the DFT-relaxed structures. \textit{This indicates that, on average, ML-FF relaxation yields lower-energy structures than DFT relaxation.} The differences in mean energies is greater than the computed bootstrap confidence interval (CI=0.95) (ID $-0.07 \pm 0.01$, OOD\_CAT $-0.13 \pm 0.09$, OOD\_ADS $-0.05 \pm 0.02 $, OOD\_BOTH $-0.13 \pm 0.01$). Notably, this is even the case for the OOD\_BOTH dataset, in which both catalysts and adsorbates are out of distribution. In each dataset, over 70\% of predicted structures are lower in energy or within 0.1 eV of the OC20 geometries. We also calculate the root-mean-square deviation of atomic positions (RMSD). Our analysis indicates that structures with lower DFT energies than OC20 correlate with higher RMSDs. This suggests that ML-FF escapes local minima to obtain qualitatively different structures with lower energies, an example of which is shown in Fig. \ref{fig:energy_dist}.

\subsection{Easy Potential Relaxations}
Does training on synthetic quadratic approximations yield similar results, which tests our hypothesis of primarily needing to train on similar minima? Our experiments for Easy Potential, presented in Table \ref{tab:Easy Potential-performance}, result in lower energy predictions than ML-FF, despite being trained on a purely synthetic potential energy surface for the initial approximation stage (CI=0.95) (ID $-0.23 \pm 0.01$, OOD\_CAT $-0.07 \pm 0.2$, OOD\_ADS $-0.18 \pm 0.01 $, OOD\_BOTH $-0.26 \pm 0.01$). In addition, Easy Potential converges in a median of 113 optimisation steps (including resolving the bond errors in the refinement stage, i.e. 50 steps for the first stage, up to 100 for the second), versus a median of 263 iterations for ML-FF. We thus obtain \textit{more accurate results with less compute time.}

To confirm that these results were not due to different DFT settings in Quantum Espresso and VASP, we also evaluate a sample of 1000 OOD\_BOTH structures using VASP, with the exact same settings used in OC20. To check our settings, we first confirmed that VASP results on the ground truth structures matched the energies provided by the OC20 dataset. We observe a distribution shift in which a higher number of structures are less stable, but still find an improvement of $-$0.13 eV on average. 

The majority of structures predicted by Easy Potential are still equivalent or lower in energy than DFT relaxations. We also carry out a study where we allow for a larger iteration budget in the refinement stage (950 versus 100 iterations in the second stage), which further improves results against VASP. Finally, we use again use the OOD\_BOTH sample relaxed via EP and continue relaxations using VASP in a hybrid approach, which results in -0.46 eV average improvements (for a median of 71 additional DFT relaxation steps) against the original VASP OC20 relaxations (median 147 DFT steps).

\begin{figure}[]
    \centering
    \includegraphics[scale=0.45]{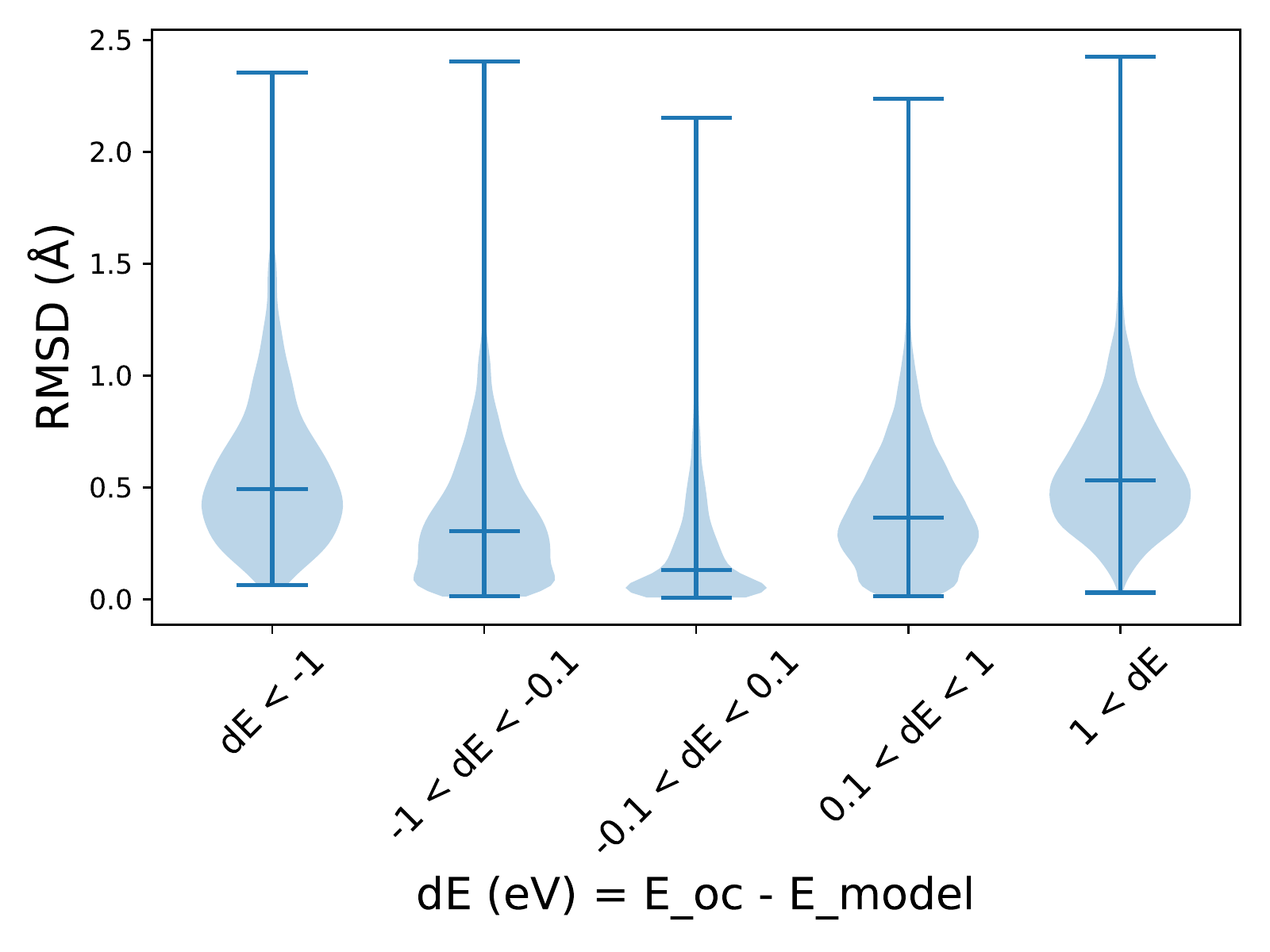}
    \caption{RMSD distribution for EP relaxations (OOD\_BOTH). Larger energy differences correspond to larger structural differences. Markers show min, median and max values respectively.}
    \label{fig:rmsd-plot}
\end{figure}

Fig.~\ref{fig:rmsd-plot} breaks down the distribution of DFT energy differences against structure differences for the OOD\_BOTH Easy Potential results in Table \ref{tab:Easy Potential-performance}. Two expected effects are observed. Structures centered around the DFT margin of error (0.1 eV) have very small average RMSD with few outliers. Mean RMSD then increases for both more stable and less stable structures.

\subsection{ML-FF Analysis}
\begin{table*}[]
    \centering
    \begin{tabular}{c|ccc|>{\columncolor{lightgray}}cc}
    Max      &   \multicolumn{3}{c}{Predicted v. OC20 DFT energies}   & Mean  & Mean\\
    iterations &   lower & within $0.1$ eV margin & higher    & ${\Delta} E$ (eV)  & RMSD (\r{A})\\
    \midrule
     100 & 31.72\% & 36.45\% & 31.83\% & 0.001 & 0.336 \\
     250 & 35.90\% & 35.57\% & 28.52\% & -0.071  & 0.282\\ 
     500  &  38.38\% & 35.37\% & 26.25\% & -0.109 & 0.363 \\
     1000  & 40.11\% & 34.07\% & 25.82\% & -0.118  & 0.335\\
     5000  & 41.09\% & 32.89\% & 26.02\% & -0.125 & 0.363 \\
    \end{tabular}
    \caption{\textbf{Effect of optimisation steps on performance.} Here we analyse the importance of the maximum number of iterations. We see that performance (for energy differences, shaded) levels out with a maximum of 1000 iterations. Ablation was carried out on OOD\_BOTH using 1000 structures.}
    \label{tab:relaxation-ablation-performance}
\end{table*}

Here we perform ablation studies to provide further insight into the performance and trade-offs of ML-FF relaxations. In Table \ref{tab:relaxation-ablation-performance}, we show that increasing the number of optimisation steps in our relaxation yields lower-energy geometries. We observe substantial improvements going from 100 to 1000 iterations, but marginal subsequent gains. Note that these results are obtained using only ML-FF, i.e. not improving results by first using the harmonic approximation for initialisation.

In Table \ref{tab:gpu_vs_cpu}, we compare runtimes with CPUs and with GPUs. Our main results were obtained using a maximum number of 1000 iterations, corresponding to a 10 minute median relaxation time on CPU and 46.5 seconds on a V100 GPU. Note that the model and relaxation procedure are deterministic, but we observe small runtime-scheduling related non-determinism on GPU platforms. Classical DFT solvers are typically run on CPUs, and CPU relaxation of Easy Potentials is fully deterministic. We expect future work to further refine execution times by additionally optimising for a specific target platform.

\begin{figure}[h]
    \centering
    \includegraphics[scale=0.5]{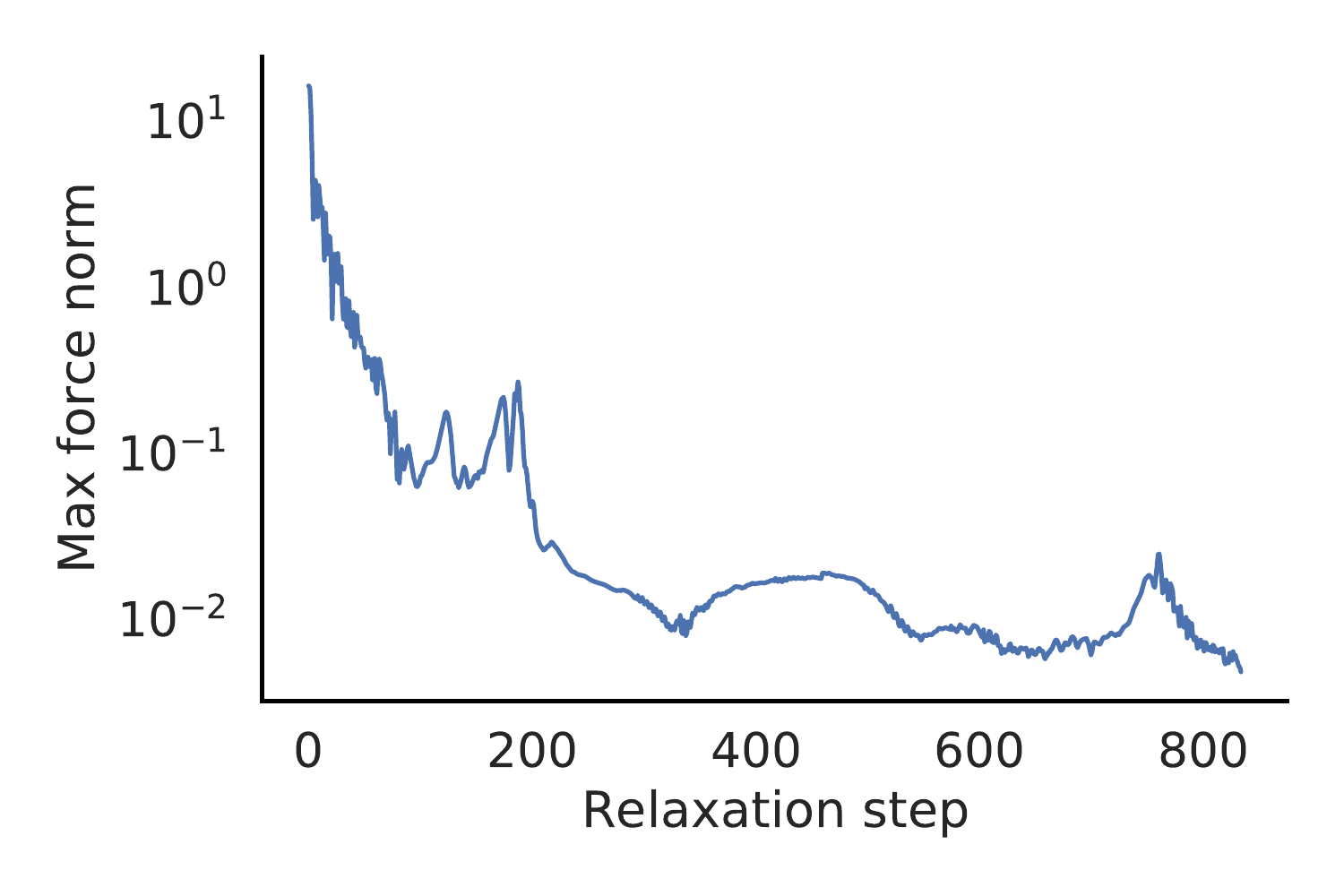}
    \caption{Example ML-FF relaxation trajectory. The maximum norm of the predicted forces is plotted against the relaxation step. Stochastic relaxation escapes local minima.}
    \label{fig:force-norm}
\end{figure}

In Fig. \ref{fig:force-norm}, we show an example ML-FF relaxation trajectory from OOD\_BOTH, illustrating how using an adaptive momentum optimiser combined with noisy force predictions allows escaping from local minima several times before converging.

\begin{table*}[]
    \centering
    \begin{tabular}{c|ccc|ccc}
    Max      &   \multicolumn{3}{c}{CPU runtime (seconds)} & \multicolumn{3}{c}{ GPU runtime (seconds)} \\
    Iter.      & Mean & Median  & 99th pctl. & Mean  & Median & 99th pctl.  \\
    \midrule
     100 & 300.0 & 274.5 & 568.1 & 17.7 & 15.5 & 26.8\\
     250 & 580.5 & 574.7 & 1316.1 & 35.1 & 36.4 & 72.1\\ 
     500  & 861.7 & 742.4 & 2479.4 & 53 & 46.3 & 136.4\\
     \rowcolor{lightgray} 1000  & 1163.2 & 752.3 & 4558.2 & 72.1 & 46.5 & 257.6 \\
     5000  & 1460.6 & 743.8 & 7202.1 & 110.0 & 48.2 & 1031.4 \\
    \end{tabular}
    \caption{\textbf{GPU vs. CPU performance for ML-FF relaxation.} V100 GPUs provide an order of magnitude speedup over CPUs. Results in the main text used a maximum of 1000 iterations (shaded), for which the median runtime is under one minute on GPU, and the 99th percentile runtime is under five minutes. }
    \label{tab:gpu_vs_cpu}
\end{table*}

\begin{figure}[h]
    \centering
    \includegraphics[scale=0.5]{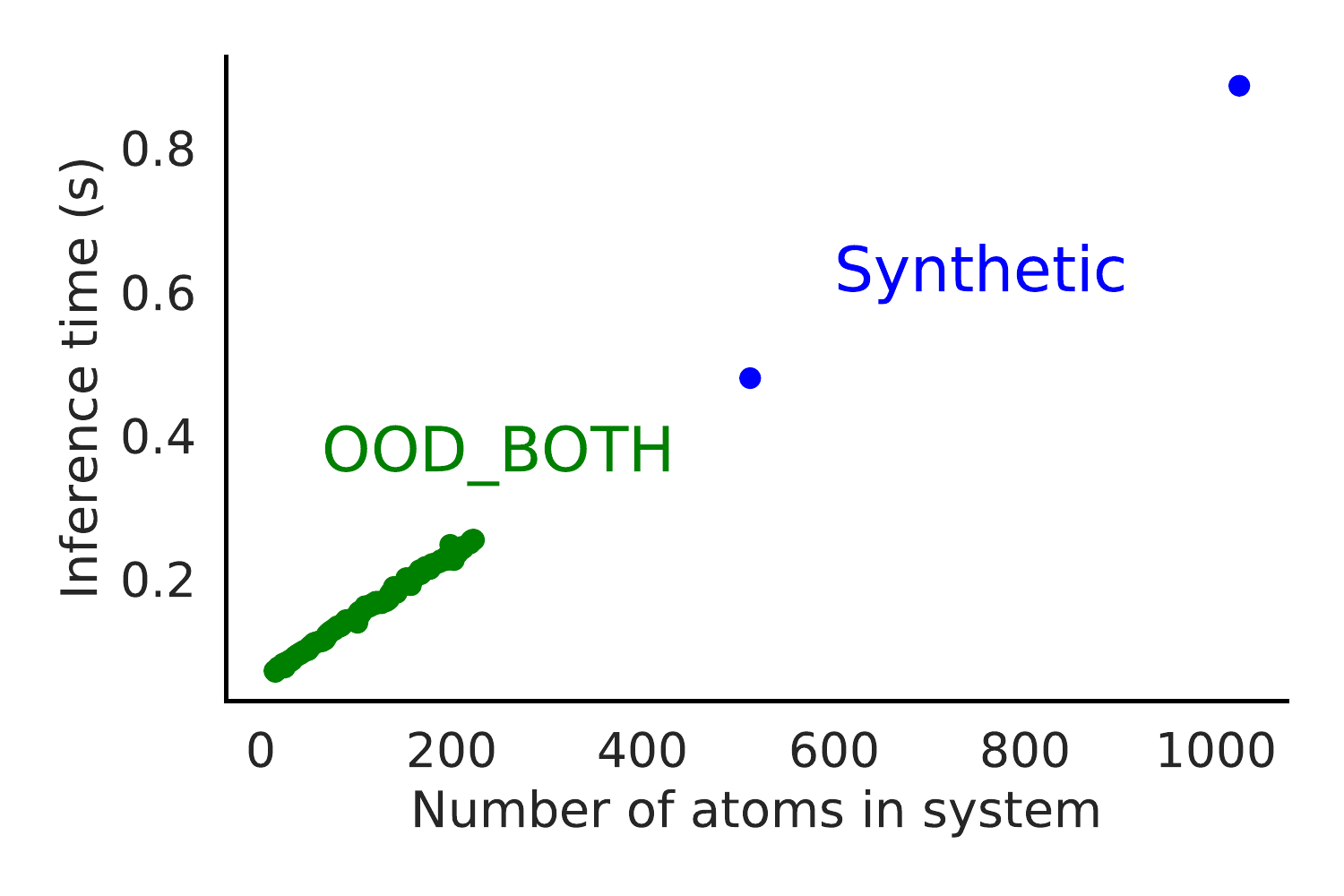}
    \caption{Scaling of inference time for different system sizes. We plot the inference time for all structures from OOD\_BOTH. Results for structures with the same number of atoms are averaged. We further show results for two larger synthetic systems. A clear linear scaling trend is observed.}
    \label{fig:inference-scaling}
\end{figure}
DFT scales cubically in system size, whereas GNS scales linearly for OC20 catalytic systems since the maximum number of interactions is capped by the cutoff radius. In Fig.~\ref{fig:inference-scaling}, we measure inference cost to predict energies and forces for structures from the OOD\_BOTH validation dataset. Executed on a single V100 GPU, we observe linear scaling against system size. The largest structures in OOD BOTH have less than 250 atoms, and we further create two synthetic systems with 512 and 1024 atoms, and with the theoretical maximum number of edges that our model admits (see Methods). We see that these large systems follow the same linear scaling trend.

\begin{figure*}[ht]
   \begin{subfigure}{.23\textwidth}
        \centering
        \includegraphics[width=\textwidth]{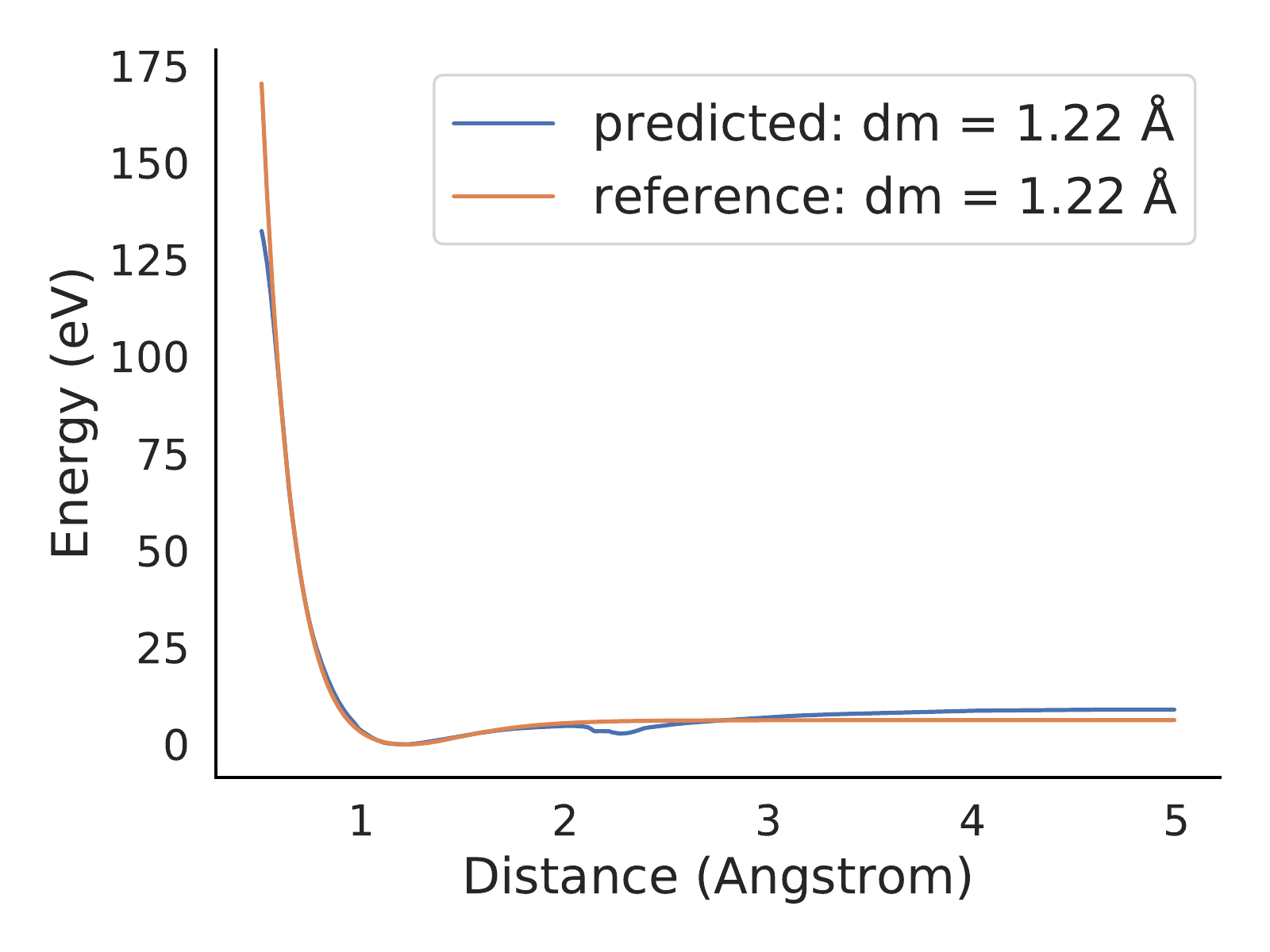}
        \caption{O-O}
    \end{subfigure}
    \begin{subfigure}{.23\textwidth}
        \centering
        \includegraphics[width=\textwidth]{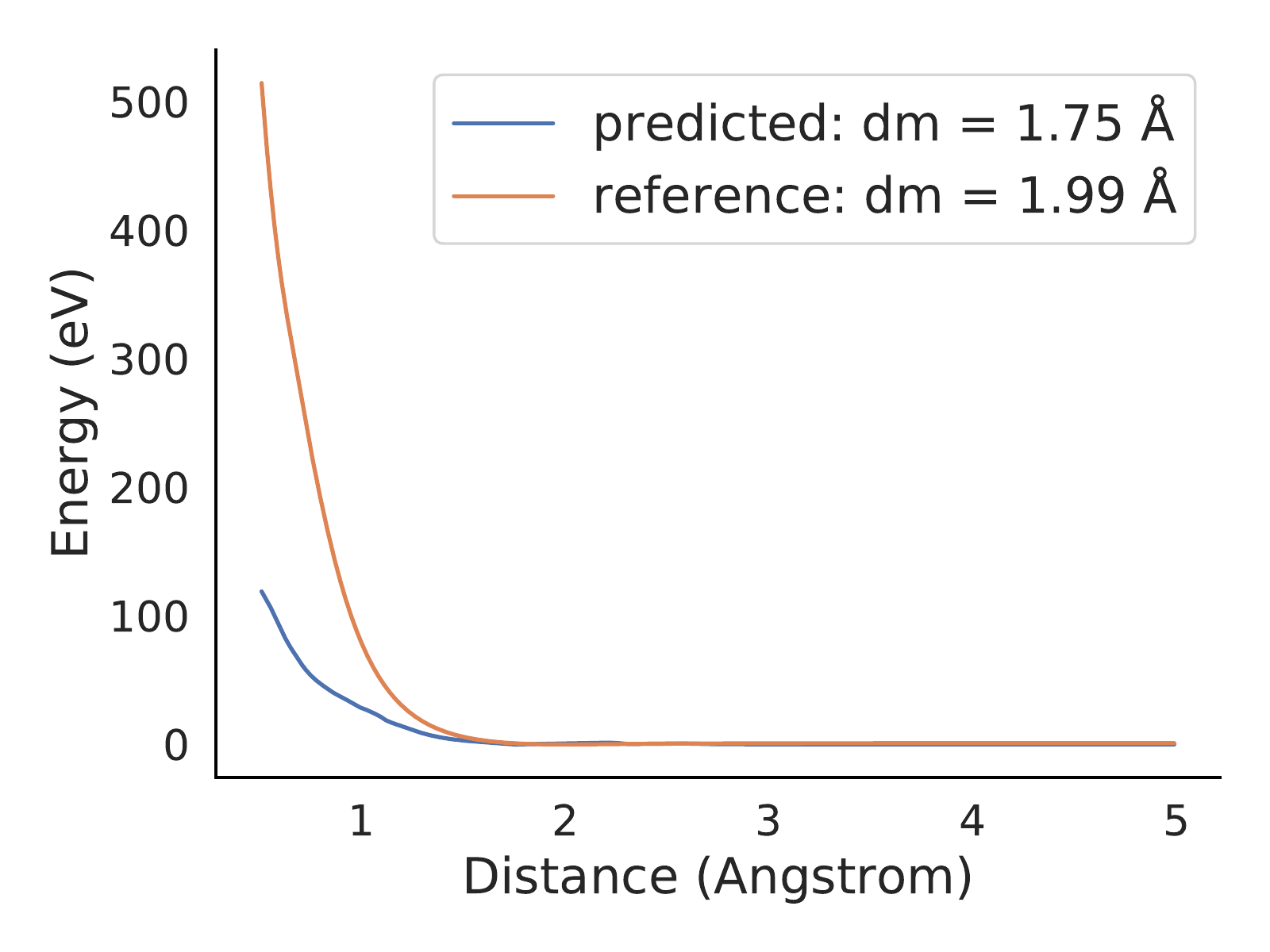}
        \caption{Cl-Cl}
    \end{subfigure}
    \begin{subfigure}{.23\textwidth}
        \centering
        \includegraphics[width=\textwidth]{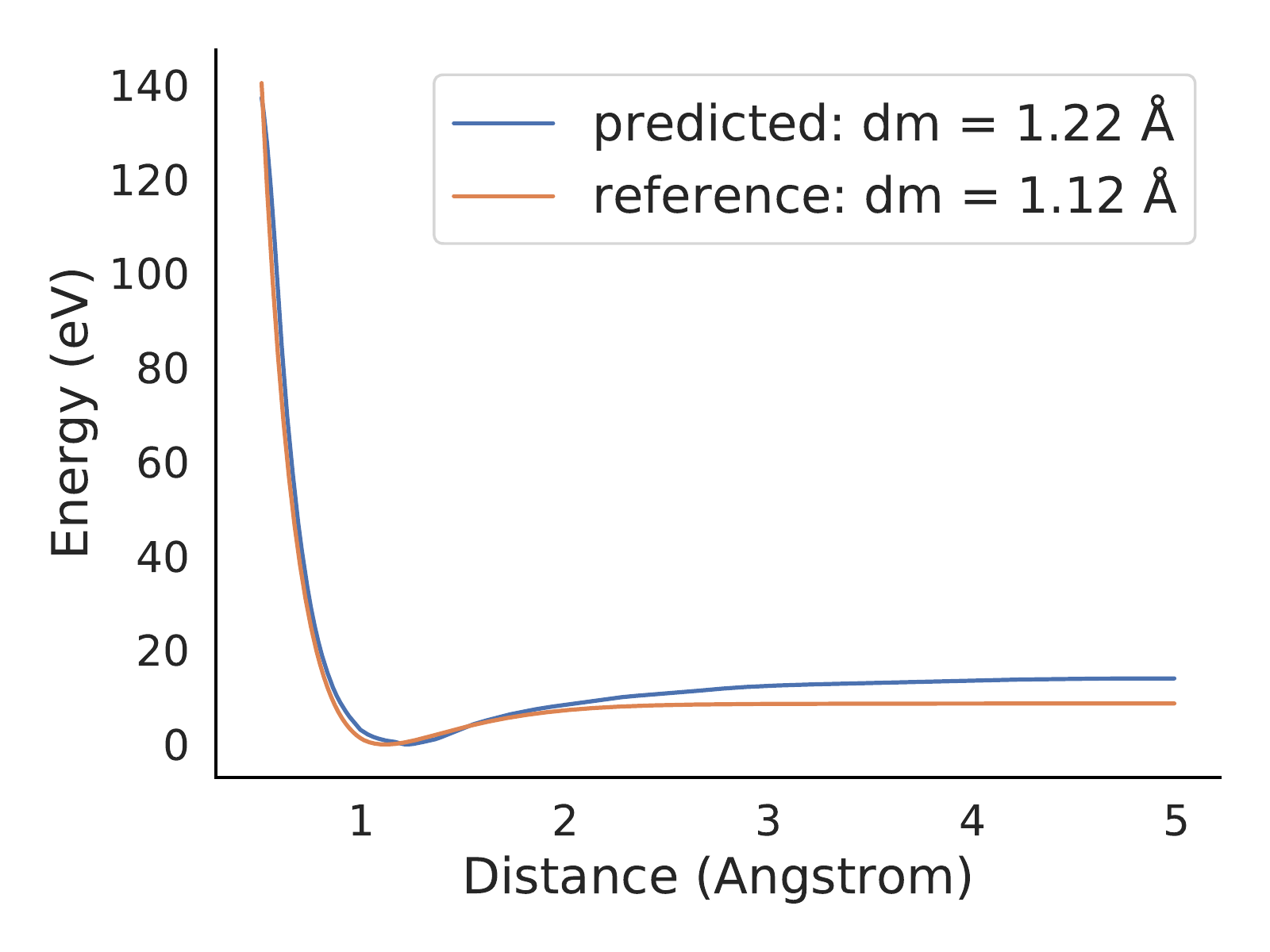}
        \caption{C-O}
    \end{subfigure}
    \begin{subfigure}{.23\textwidth}
        \centering
        \includegraphics[width=\textwidth]{assets/bond_length_curves/Cl-Cl.pdf}
        \caption{P-P}
    \end{subfigure}
    \vskip\baselineskip
    \begin{subfigure}{.23\textwidth}
        \centering
        \includegraphics[width=\textwidth]{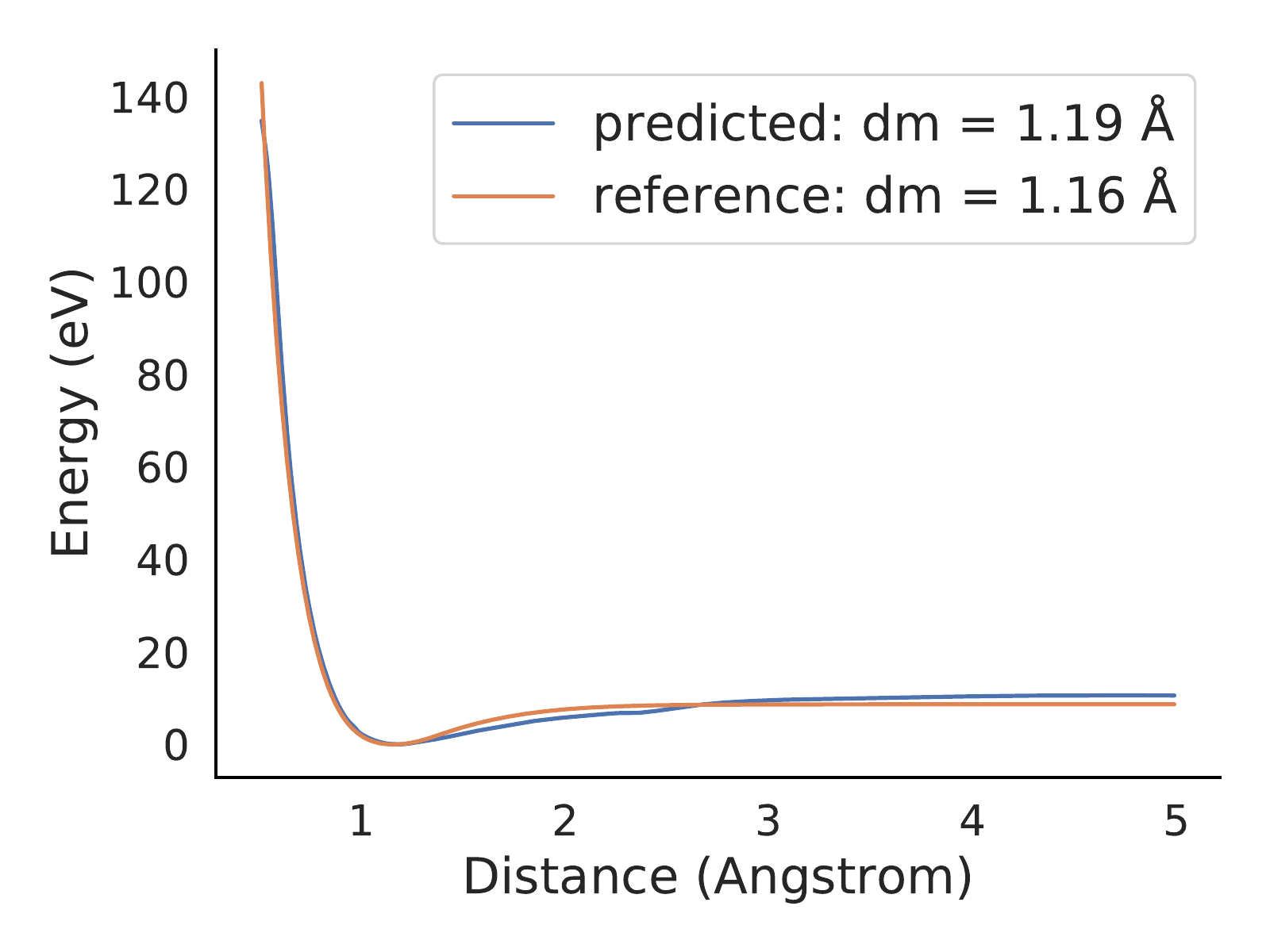}
        \caption{N-O}
    \end{subfigure}
    \begin{subfigure}{.23\textwidth}
        \centering
        \includegraphics[width=\textwidth]{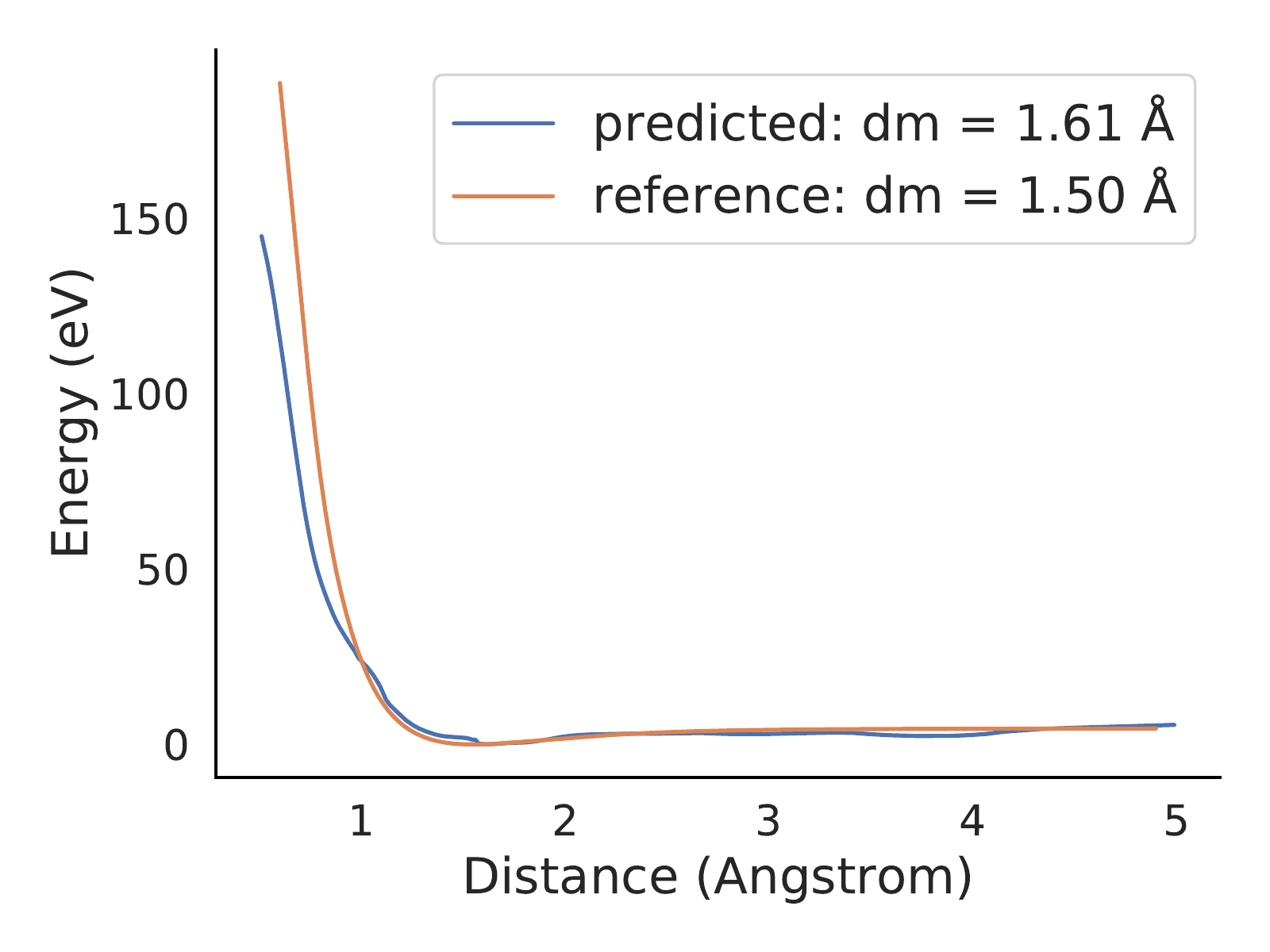}
        \caption{O-Si}
    \end{subfigure}
    \begin{subfigure}{.23\textwidth}
        \centering
        \includegraphics[width=\textwidth]{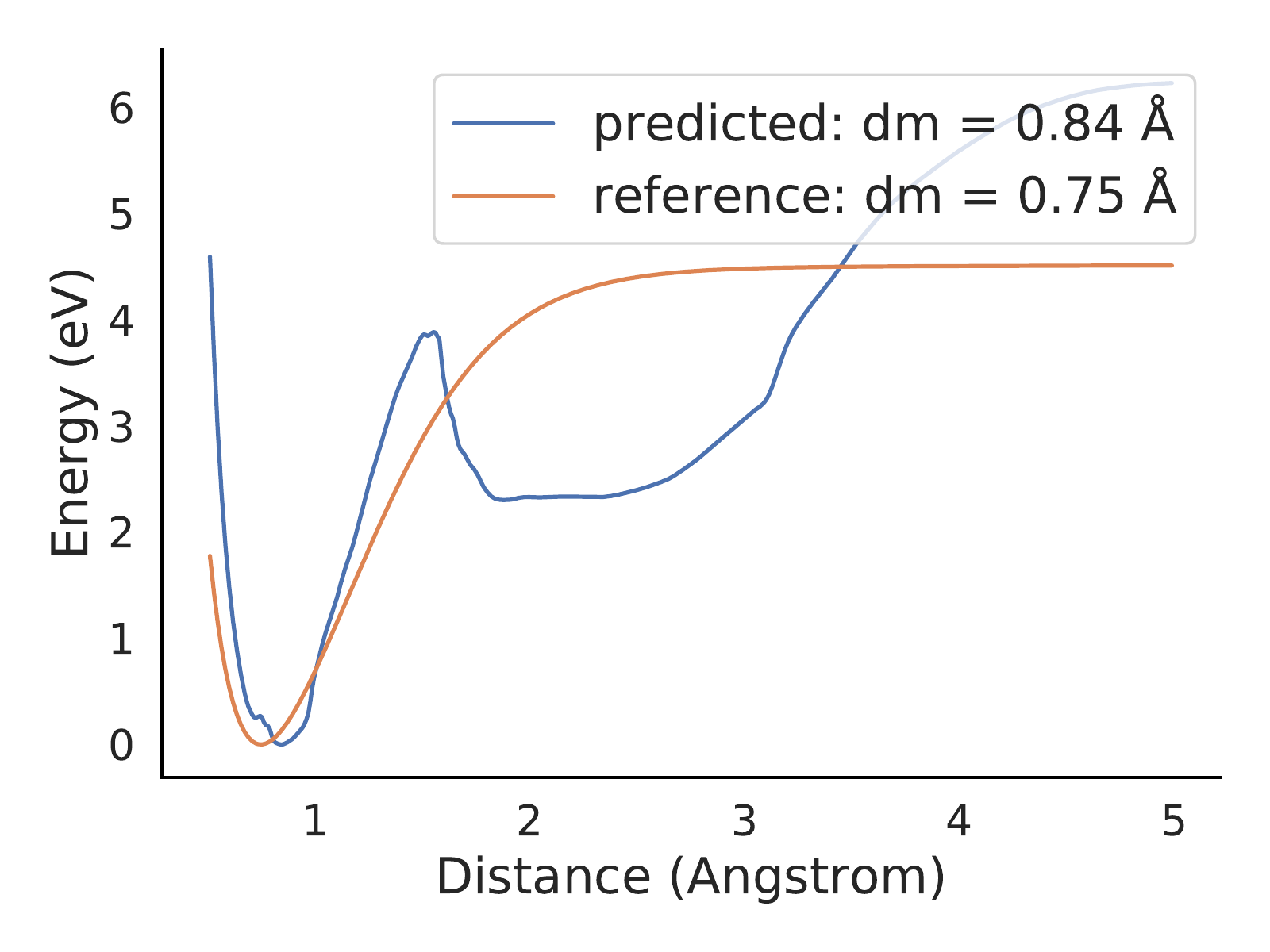}
         \caption{H-H}
    \end{subfigure}
    \begin{subfigure}{.23\textwidth}
        \centering
        \includegraphics[width=\textwidth]{assets/bond_length_curves/Cl-Cl.pdf}
        \caption{Se-Se}
    \end{subfigure}
      \caption{\textbf{Potential energy of atom pairs.} Using Quantum Espresso (reference) \cite{Giannozzi_2009} and our ML-FF model (predicted), we compute the potential energy of several atom pairs as a function of the distance separating the nuclei.}
      \label{fig:energy_bond_dist}
\end{figure*}

\subsection{Bond Length Analysis}
We further validate ML-FF's ability to predict minima by evaluating bond lengths. To do so we predict the potential energy of atom pairs at various bond distances. The bond length is the distance that minimizes the energy. We compare ML-FF predictions with the ground truth computed with Quantum Espresso. The results are illustrated in Fig. \ref{fig:energy_bond_dist}: the predicted distance is within 10\% of the target for all the pairs we sampled. Furthermore, our model and the ground truth tend to show similar behavior near the energy minimum, in many cases displaying non-parabolic behaviour. We observe more difficulties on some pairs (e.g. Cl-Cl, P-P, Se-Se) which is likely due to a limited number of these bonds in the training data.

\section{Discussion}
The key finding of our experiments is that learned interatomic potentials are surprisingly accurate at identifying low energy wells of target DFT PESs, even if force predictions are noisy or the model is trained on a simple harmonic potential. The latter are similar in spirit to gradient fields developed for ML conformer generation \cite{shi2021learning, luo2021predicting}, which are trained on minimum-energy structures rather than actual forces. We speculate that DFT packages like VASP struggle to escape local minima due to using classical optimisers and/or non-noisy ground truth forces. A detailed investigation of the importance of these two factors is left for future work. The main impact of our findings is that learned force fields are demonstrably useful for practical calculations today.

Using ML-FFs and Easy Potentials opens up the possibility of relaxing large systems of tens of thousands of atoms due to their fast execution and linear scaling. We imagine Easy Potentials being used in collaboration with classical methods, for example to obtain an initial fast result within a low energy well, which can then be further refined. As shown when combining EP with VASP in table \ref{tab:Easy Potential-performance}, this combines the guarantees of DFT while reducing the number of steps required while also improving results.

Lastly, our results imply that focusing on force and structure distance errors on Open Catalyst and similar datasets may not accurately capture the progress in learned force fields. Instead, predicted structures should be evaluated with single-point DFT calculations. Practitioners investigating other approaches to learned potentials may be similarly surprised to find that, by incorporating harmonic potentials or novel optimisation strategies for their relaxations, their force fields may already be sufficiently accurate for ground state catalyst discovery.

\section{Methods}
\subsection{Model details}\label{model}
Our model is a variant of the Graph Net Simulator \cite{pmlr-v119-sanchez-gonzalez20a} with the modifications described by Godwin et al. \cite{godwin2022simple} for molecular property prediction. Below, we provide a high level summary of the model and configurations. Please refer to the original papers for comprehensive explanations.

GNS consists of an encoder that processes input data into a latent space, a processor that executes message passing and updates the latent representation, and decoders that interpret the resulting node embeddings to per-node and graph-level predictions. 
In the encoder, we use the set of atoms to construct a directed graph $G(V, E)$ with featurized sets of vertices $V =\{v_1, v_2, \dots, v_{\left| S \right|}\}$ and edges $E = \{ e_{i,j} \}_{i,j}$ where an edge between $v_i$ and $v_j$ is added if their distance is less than a connectivity radius of 6 \AA, for a maximum of 20 edges per atom determined via a k-d tree for nearest neighbours. Each edge $e_{i,j}$ is represented with the displacement vector $p_j - p_i$ and the Euclidean norm as features. Distances are then further featurised by computing Gaussian radial basis functions (mean $\mu=0$, standard deviation $\sigma=0.5$ \AA). For each atom ($a_i$, $p_i$) a vertex $v_i$ is represented through a learned embedding of the atom type $a_i$. For OC20, we also utilise another input feature, a ``tag'' for each atom that indicates whether it is part of the slab, surface, or catalyst, by creating a separate learned embedding. Embedding sizes were set to 512, and all multi-layer perceptrons (MLPs) had size 1024 and used shifted softplus activations.

Following the input graph construction, the processor iteratively updates node and edge features by applying Interaction Networks \cite{Battaglia2016InteractionNF}. Each application represents one message passing step with a different set of weights. We repeat this for a configurable number of message passing steps, and then apply the entire processor multiple times sequentially while re-using weights. During training we average the output from each application. At inference time we only use the output from the final step. We used 10 repeated applications of 5 message passing layers.

Finally, after message passing, forces are decoded by applying an MLP to each node feature. Energy predictions are decoded by applying an MLP to each node feature, aggregating the node-level results using a sum aggregation function, and finally applying another MLP to predict the total potential potential energy of the system. Note that energy predictions are not strictly required for relaxations if not differentiating them. We primarily use them for assessing the model's internal evaluation consistency during relaxation (i.e. for analysis without DFT).

GNS is implemented in Jax \cite{jax2018github} using Haiku \cite{haiku2020github} for layer implementations, Jraph \cite{jraph2020github} for GNN components, Optax \cite{optax2020github} for optimisation, and Automap for distributed training compilation \cite{automap21}. Harmonic potentials are trained for 250 000 gradient steps on the OC20 ISRS dataset using Gaussian noise around positions which were interpolated between initial and finial relaxed structures ($\mu=0, \sigma=0.5$). Force predictions on S2EF for ML-FF are then trained for up to 2 million gradient steps with early stopping for model selection. The two step procedure follows the methodology described by Zaidi et al. \cite{zaidi22}.

All training is performed on a cluster of 16 TPU v4 devices with 32 GB of RAM. Graphs are dynamically batched with up to 10 graphs, 1024 vertices and 12800 edges in a single batch (i.e. the batch is padded to these dimensions once one of them is reached, since input structures have varying dimensions). Parameters are updated using a mean squared error loss and the Adam optimiser \cite{Kingma2015AdamAM} ($\beta_1=0.9$,  $\beta_2=0.95$) with a warm up ($10^5$ warm up steps, $10^{-5}$ warmup start learning rate and $10^{-4}$ warmup max learning rate), and a cosine decay schedule ($5 \cdot 10^6$ cosine cycle length with $10^{-6}$ min learning rate). Finally, parameters are smoothed using an exponentially moving average decay with a decay parameter of $0.9999$.

\subsection{Relaxation settings}
Relaxations are implemented using the Adamw \cite{loshchilov2018fixing} optimiser from the Optax library \cite{optax2020github} (decay parameters $b_1=0.9,b_2=0.99$). A learning rate of $0.1$ is used to update structures. Relaxations are run with two stopping conditions (stopped when either is reached):
\begin{itemize}
    \item A maximum number of relaxation steps (ML-FF: 1000, Easy Potential/ML-FF two-stage procedure: 50 Easy Potential, 100 ML-FF).
    \item The Frobenius norm of the predicted forces per-atom below a threshold hyper-parameter for all atoms (ML-FF: $0.005$, Easy Potential/ML-FF two-stage: $0.03$).
\end{itemize}
As gradients for Adamw, we use the negative direct per-atom force predictions rather than explicitly computing the derivative of the predicted energy with respect to positions. As discussed in the Model Details section, atom graphs are constructed based on radial distance cutoffs. The relaxed atomic positions change inter-atomic distances. We hence recompute edges after updating positions by re-executing a nearest neighbour lookup within the distance cutoff.

Since relaxations are implemented in JAX, we in practice use jit compilation to XLA which requires static shapes. Graphs change shape throughout relaxations, so we compile ahead-of-time to an initial power-of-two for edge counts and pad graphs to this size, and only recompile dynamically as needed to the next power of two to minimize the total number of recompilations throughout a relaxation. For large scale inference, pre-compiling a range of sizes and matching each molecule to the next power-of-two could avoid any runtime compilation and optimize throughput. Relaxations are measured on both CPUs and NVIDIA V100 GPUs with 16 GB RAM.

\subsection{Dataset preparation}
We used the OC20 dataset as provided in https://github.com/Open-Catalyst-Project/ocp. We conducted some data cleaning, removing 580 structures where the surface crossed the unit cell $z$-axis. This was observed after training, and future models could be adjusted to also handle a periodic $z$-axis if necessary.

\subsection{DFT settings}
We faithfully reproduce the settings described in the Open Catalyst paper ~\cite{Chanussot2020TheOC}, which utilises VASP, while we use the open source Quantum Espresso package (apart from a small sample of VASP validations). We use the RPBE functional and computed the $k$-points for each structure using the same formula as ~\cite{Chanussot2020TheOC}:

\[ K = \left( \left\lfloor \frac{m}{\mathbf{a}_1} \right\rceil, \left\lfloor \frac{m}{\mathbf{a}_2} \right\rceil , 1 \right) \]

Here $m=40$ and $\mathbf{a}_1$, $\mathbf{a}_2$ are the dimensions in the first and second axis of the unit cell of the structure. Since a significant part of our structures are metallic we used a Marzari-Vanderbilt smearing with a Gaussian spreading of $0.2$. We do not observe a significant energy difference for the insulators of the dataset with or without smearing. Therefore, for practical reasons, we used smearing for all DFT computations. Moreover, due to computational limitations we do not use the precise pseudo-potentials in \cite{Prandini_2018}, but rather choose the most efficient ones. Finally, to improve DFT convergence rates, we use a local density-dependent Thomas-Fermi screening for the mixing mode, with beta equal to $0.7$. Each DFT computation was launched on a Google Cloud Platform (GCP) machine with 62 CPUs and 248 GB of RAM.

We use VASP for a small sample of structures for validation. Our VASP settings exactly reproduce the OC20 results when using the configurations specified in the paper \cite{Chanussot2020TheOC}, together with the corresponding preprocessing scripts that were released with the datasets. The jobflow software was used for automating VASP calculations \cite{jobflow}.

\section{Data availability}
The predicted structures for ML-FF and Easy Potential will be released upon acceptance of this manuscript.

\section{Code availability}
Model code, trained checkpoints, as well as the relaxation implementation will be released upon acceptance of this manuscript.

\section{Author contributions}
Jonathan Godwin and Michael Schaarschmidt carried out the original research into learned potentials, including model design and implementation. Morgane Rivière designed and implemented the DFT and relaxation evaluation infrastructure. Alex M. Ganose carried out structure validations and advised on experiment design for evaluating relaxed structures. James S. Spencer implemented the first DFT validation prototype and advised on the configuration of Quantum Espresso, as well as contributed to the interpretation of results, writing and analysis. Alexander L. Gaunt and James Kirkpatrick advised on computational chemistry and relaxation evaluation. Simon Axelrod advised on computational chemistry and contributed to writing. Peter W. Battaglia advised on initial model design for learned potentials as well as on the overall project structure and delivery.


\section{Acknowledgments}
We thank Alvaro Sanchez-Gonzalez and Ivo Danihelka for their helpful suggestions and comments on earlier versions of this manuscript. We further are grateful to Joe Stanton for helping us scale DFT experiments on Google Cloud, as well as to Alison Reid for guiding the open sourcing process. A.M.G. was supported by EPSRC Fellowship EP/T033231/1.
\bibliography{main}
\end{document}